\documentclass[a4paper,12pt]{article}
\usepackage{amsfonts,amsthm,amsmath,amssymb,graphicx,hyperref,youngtab}
\usepackage[bulletsep]{collref}

\def\Tr{{\rm Tr }}

\def\G{\Gamma}

\newcommand{\be}{\begin{equation}}
\newcommand{\bea}{\begin{eqnarray}}
\newcommand{\ee}{\end{equation}}
\newcommand{\eea}{\end{eqnarray}}

\begin{document}

\makeatletter
\@addtoreset{equation}{section}
\makeatother
\renewcommand{\theequation}{\thesection.\arabic{equation}}

\rightline{WITS-CTP-106}
\vspace{1.8truecm}

\vspace{15pt}


{\LARGE{
\centerline{   \bf From Schurs to Giants in ABJ(M)}
\centerline {\bf }
}}

\vskip.9cm

\thispagestyle{empty} \centerline{
    {\large \bf  Pawe{\l} Caputa${}^{a,} $\footnote{ {\tt pawel.caputa@wits.ac.za}}  and Badr Awad Elseid Mohammed${}^{a,b} $\footnote{ {\tt bmohamme@ictp.it}}}
                                                    }

\vspace{.8cm}
\centerline{{\it ${}^a$ National Institute for Theoretical Physics}}
\centerline{{\it Department of Physics and Centre for Theoretical Physics }}
\centerline{{\it University of the Witwatersrand, Wits, 2050, } }
\centerline{{\it South Africa } }
\vspace{.4cm}
\centerline{{\it ${}^b$ Department of Physics}}
\centerline{{\it Sudan University of Science and Technology }}
\centerline{{\it 407, Sudan} }

\vspace{1.4truecm}

\thispagestyle{empty}

\centerline{\bf Abstract}
\vskip.5cm
In this work we consider various correlators with Schur polynomials in ABJ(M) models that on the dual gravity side should correspond to processes involving giant gravitons. Our analysis imposes several constraints on the physics of the probe branes on $AdS_4\times \mathbb{CP}^3$ as well as sheds more light on giant graviton solutions in this background with additional NS B-field. Our main tool is a formula that we derive for extremal n-point functions of the single trace chiral primary operators in the free field theory limit. The formula expresses the correlators in terms of the two-point function of Schur polynomials labeled by hook diagrams and is valid for a large class of gauge theories. In particular, in $\mathcal{N}=4$ SYM, it proves the conjecture of \cite{Beisert:2002bb}.

\setcounter{page}{0}
\setcounter{tocdepth}{2}

\newpage

 \noindent\rule\textwidth{.1pt}

 \tableofcontents

 \vskip 2em 
 \noindent\rule\textwidth{.1pt}

\setcounter{footnote}{0}

\linespread{1.1}
\parskip 4pt

{}~
{}~

\section{Introduction and summary}
Since the first concrete model for holography \cite{Maldacena:1997re,Witten:1998qj}, decoding geometry from the dual field theory has been one of the most important steps in understanding gauge/gravity correspondence. Successful examples end up in the holographic dictionary that consists of a set of gauge invariant operators in field theory and corresponding objects in gravity that share the same physics. Probe D-branes (or M-branes) in  $AdS$ backgrounds known as giant gravitons \cite{McGreevy:2000cw,Grisaru:2000zn,Hashimoto:2000zp} and their dual Schur polynomial operators \cite{Corley:2001zk} contribute to an important chapter in this book.  

\medskip
Last years of progress in describing M2-branes provided a new holographic platform where the fidelity of the translation from Schurs to giants can be tested.  Three-dimensional ABJ(M) models \cite{Aharony:2008ug,Aharony:2008gk} are conjectured to capture the physics of M-branes in $AdS_4\times S^7$ and D-branes in type IIA theory on $AdS_4\times\mathbb{CP}^3$. Such a rich structure of the duality allows for a zoo of giant graviton solutions as well as determinant-like operators that can be constructed and tested against each other.

\medskip
We choose the gauge theory as a framework of our analysis and explore the Schur polynomial operators in ABJ constructed in \cite{Dey:2011ea}. More precisely we compute various correlators that, on the gravity side should involve giant gravitons in $AdS_4\times \mathbb{CP}^3$ with background NS B-field. In particular radiation of closed strings from giants, splitting and joining of open strings on excited gravitons and three point correlators of two heavy giants and one light graviton. After detailed analysis we put forward a prediction for the ABJ giant graviton solution that remains to be constructed and should generalize \cite{Giovannoni:2011pn}. 

\medskip
The key to evaluation of amplitudes with open strings attached to giants are the extremal correlators of half-BPS chiral primary operators (CPO). In $\mathcal{N}=4$ SYM their exact two, three and four point functions were first computed in \cite{Kristjansen:2002bb} using matrix model and Schurs, and form of a general n-point function was conjectured in \cite{Beisert:2002bb,Okuyama:2002zn}. Using Schur polynomials\footnote{See also \cite{Corley:2002mj} for application of Schurs in this context.} we are able to give a constructive proof of this conjecture and generalize it to a larger class of field theories where CPOs can be expanded in a basis of Schurs. With our result we then derive extremal n-point correlators in ABJ(M) in the free field theory limit but to all orders in gauge group ranks $N$ and $M$ and use them to evaluate relevant correlators with giants.

\medskip
A summarizing list of our results is the following:
\begin{itemize}
\item Extremal n-point correlators in ABJ(M) are expressed in terms of Meijer G-function and can be written in a similar form as their $\mathcal{N}=4$ SYM counterparts
\item Antisymmetric Schur polynomials as well as the amplitudes for radiation and string joining only exist when the number of boxes or giants momentum is smaller than the smaller rank. In this work we choose it to be $M$
\item Open strings in ABJ come in two species, as $N\times N$ or $M\times M$ words that we call N- or M-strings. Each family can be attached to a giant in a particular way
\item Maximal giant gravitons do not radiate
\item Leading order amplitudes are parity invariant but subleading corrections break parity
\item Only probes dual to Schurs with hooks have a non-zero overlap with point-like gravitons
\item Three point functions of two heavy Schurs and one light graviton in ABJ exhibit an interesting doubling structure
\end{itemize}
Discussion of these results and the emerging picture for the dual graviton on $\mathbb{CP}^3$ with NS B-field will be provided in the main text.

\medskip
This article is organized as follows. We start with a brief review of the ABJ(M) models and Schur polynomials in these theories. In section \ref{ABJMSchurs} we compute extremal n-point correlators of the single trace chiral primary operators in ABJ(M) in the free field theory limit. We then use these results in sections \ref{OpenStrings} and \ref{MoreCorrelators} to study correlators with giant gravitons from ABJ(M) perspective and describe the emerging prediction for the giant graviton dual to Schurs in ABJ. Appendix \ref{CorrelatorsN4} contains a proof of conjecture \cite{Beisert:2002bb} and \ref{Hooks},\ref{Proof},\ref{Details} additional details of our derivations.

\section{Brief review of the ABJ(M) models}\label{ABJ(M)}
Over the past few years there has been a great progress in understanding dynamics of M2 branes \cite{Bagger:2012jb}. One of the most important developments in this direction is the new $AdS_4/CFT_3$ duality proposed by Aharony, Jafferis, Bergman and Maldacena (ABJM)\cite{Aharony:2008ug}. The conjecture identifies a three dimensional $(\mathbb{R}^{1,2})$, $\mathcal{N}=6$ superconformal Chern-Simons-matter (CSm) gauge theory with $U(N)_k\times U(N)_{-k}$ group ($k$ is the CS level) and matter in the bi-fundamental representation (the ABJM model), as the world-volume description of $N$ M2-branes on $\mathbb{C}^{4}/\mathbb{Z}_{k}$ orbifold singularity. Soon after the original proposal, three of the authors \cite{Aharony:2008gk}, generalized the duality to CSm theory with $U(N)_{k}\times U(M)_{-k}$ gauge group (the ABJ model). In M-theory this corresponds to $min(M,N)$ M2-branes moving freely on $\mathbb{C}^{4}/\mathbb{Z}_{k}$ and $|M-N|$ fractional M2-branes stuck on the orbifold singularity. In addition there are $|M-N|$ units of discrete torsion for the background 4-form. 

\medskip
Details of the gauge theories were worked out in \cite{Benna:2008zy}. The field content of ABJ(M) is given by two gauge fields, four Weyl spinors and four complex scalars $(A_1,A_2,B^{\dagger}_1,B^{\dagger}_2)$. Actions for ABJ and ABJM have precisely the same form and the only difference is that fields in the former model are rectangular $N\times M$ matrices, so ABJ explicitly breaks parity\footnote{Parity in this context exchanges two gauge groups}. In perturbation theory, CS parameter $k$ plays a role of the coupling constant (interactions are suppressed as $1/k$). Therefore, two different ranks make it possible to define two 't Hooft couplings $\lambda=N/k$, $\lambda'=M/k$, and two 't Hooft limits $k,N,M\to \infty$ while $\lambda,\lambda'$-fixed. Moreover, in order for the theory to be consistent on the quantum level one should also require $|M-N|\leq k$.

 \medskip
The new duality shares various features with the old $AdS_5/CFT_4$, but in many aspects it is much richer. One interesting novelty is that depending on the mutual range of parameters $k$ and $N$, dual theory in the bulk is either M-theory or type IIA supergravity. More precisely, in the 't Hooft limit and $N^{1/5}\ll k\ll N$ M-theory reduces to the type IIA  on $AdS_4\times \mathbb{CP}^3$ with $N$ units of RR four-form flux through $AdS_4$ and RR two-form flux $F^{(2)}\sim k$ through $\mathbb{CP}^1\subset \mathbb{CP}^3$. Additional modification for the 't Hooft limits in ABJ is the non-trivial discrete holonomy of the NS field B2 along $\mathbb{CP}^1\subset \mathbb{CP}^3$.\\  This range dependence of the duality is encoded in the dictionary between parameters on both sides 
\begin{equation}
g_s\sim \left(\frac{N}{k^{5}}\right)^{1/4}=\frac{\lambda^{5/4}}{N},\qquad \frac{R^2}{\alpha'}=4\pi \sqrt{2\lambda},
\end{equation} 
where $g_s$ is the string coupling and $R$ the radius\footnote{This second relation is modified at higher loops \cite{Bergman:2009zh}} of $\mathbb{CP}^3$ and twice the radius of $AdS_4$. 

 \medskip
The $AdS_4/CFT_3$ conjecture has already passed various non-trivial tests \cite{Bagger:2012jb}. Again, planar integrability of the models proved very useful in understanding spectrum related details of the correspondence\footnote{For review see \cite{Klose:2010ki} and further references therein}. However, a great new feature that appeared for the first time in ABJ(M) is that the correspondence can be tested at the level of quantum gravity. Namely, using localization \cite{Pestun:2007rz}, the free energy has been computed at any $\lambda$ and finite $N$ \cite{Drukker:2010nc} (see also \cite{Marino:2011nm} for pedagogical review), and its tests are at present very active area of research (see e.g. \cite{Fuji:2011km, Hanada:2012si,Bhattacharyya:2012ye}). One might hope that extending these techniques further will one day allow for comparing not only the free energy but also other observables including correlators of less supersymmetric gauge invariant operators.

\section{Schurs and giants in $AdS_4/CFT_3$ }\label{SchursABJ(M)}

 Schur polynomials provide a natural and very useful basis for half-BPS operators with large R-charge \cite{Corley:2001zk}. In $\mathcal{N}=4$ SYM, they are defined as
\begin{equation}
\chi_T(Z)=\frac{1}{n!}\sum_{\sigma\in S_n}\chi_{T}(\sigma)\,Z^{i_1}_{i_{\sigma(1)}}\cdots Z^{i_n}_{i_{\sigma(n)}},\label{SN4}
\end{equation}  
where $Z$ is a scalar in the adjoint of $U(N)$, $T$ is a Young diagram with $n$ boxes and $\chi_{T}(\sigma)$ is the character of permutation $\sigma\in S_n$ in representation $T$. Repeated indices $i$ are summed over so Schurs are linear combinations of multi-trace operators. Recall that single and multi trace operators neither are orthogonal and form a useful basis nor are good observables when the R-charge grows as $O(N)$. On the contrary, two and higher point correlators of Schurs are completely determined by representations of the symmetric group.

\medskip
Schurs \eqref{SN4} not only furnish a convenient basis in the half-BPS sector but also have a clear holographic duals in terms of D3-branes wrapping three-cycles in $AdS_5\times S^5$ \cite{McGreevy:2000cw,Grisaru:2000zn,Hashimoto:2000zp}. The dictionary states that operators labelled by a single column (completely antisymmetric representation) or a single row (completely symmetric), are gauge theory duals of probe D3 branes wrapping $S^3\subset S^5$ or $S^3\subset AdS_5$ respectively, both stabilized by rotation on $S^5$. The probes are customarily dubbed to giant gravitons (in $S^5$) or dual giants (in $AdS_5$). These identifications are based on quantitative arguments like bound on the angular momentum, but also quantitatively by free fermions wave functions \cite{Berenstein:2004kk} or recent subtle computations with semiclassical three point functions \cite{Bissi:2011dc,Caputa:2012yj,Lin:2012ey}.  For further literature on Schur polynomials in $AdS_5/CFT_4$ see \cite{Corley:2001zk,Corley:2002mj,Balasubramanian:2001nh}.

\medskip
In ABJ, complex scalars $A$ are $N\times M$ matrices that transform in $(N,\overline{M})$ and $B^{\dagger}$ are $M\times N$ matrices transforming in $(\overline{M},N)$. Both sets have the conformal dimension equal to the R-charge that is 1/2. The simplest half-BPS operators that one can construct are then multi-traces built of pairs $AB^{\dagger}$ or $BA^{\dagger}$ \footnote{The R-charge of these pairs is equal to their conformal dimension that is 1.}\footnote{Note that the pair effectively transforms in the adjoint of $U(N)_k$. Equivalently one could choose the combination $B^{\dagger}A$ in the adjoint of $U(M)_{-k}$. }. The distinction between $A_1$ and $A_2$ (or $B_1$ and $B_2$) will be important in construction of operators dual to giants with strings attached, but for now $A$ $(B)$ stands for any of the two. In a complete analogy with \eqref{SN4}, Schurs in ABJ(M) are then defined as \cite{Dey:2011ea,Chakrabortty:2011fd}
\begin{equation}
\chi_{T}(AB^{\dagger})=\frac{1}{n!}\sum_{\sigma \in S_n}\chi_{T}(\sigma)\,(AB^{\dagger})^{i_1}_{i_{\sigma(1)}}\ldots (AB^{\dagger})^{i_n}_{i_{\sigma(n)}}.\label{ABJS}
\end{equation}

As shown in \cite{Dey:2011ea}, ABJ(M) Schurs form an orthogonal basis and their two point functions are diagonal\footnote{In this work we always omit the space time dependence of the correlators that, as in any CFT, can be easily restored at any stage of calculations.}. For two Young diagrams $R_1$ and $R_2$ the two point function of corresponding Schur polynomial operators in $AB^{\dagger}$ is given by
\begin{equation}
\langle \chi_{R_1}(AB^{\dagger})\chi_{R_2}(BA^{\dagger})\rangle=\delta_{R_1 R_2}\,f_{R_2}(N) f_{R_2}(M),\label{2ptSchurs}
\end{equation}
where $f_{R}(N)$ and $f_{R}(M)$ are products over the weights of each box in $R_i$.\footnote{Appendix \ref{Proof} contains the proof of this formula using projection operators.}

What makes Schur polynomials particularly useful in gauge theory computations is the Littlewood-Richardson fusion rule \cite{Fulton} that expresses a product of two Schurs labeled by diagrams with $k_1$ and $k_2$ boxes by a linear combination of Schurs with diagrams of size $k_1+k_2$. In ABJ(M) the rule becomes
\begin{equation}
\chi_{R_{k_1}}(AB^{\dagger})\chi_{R_{k_2}}(AB^{\dagger})=\sum_{R_{k_1+k_2}}g(R_{k_1},R_{k_2};R_{k_1+k_2})\,\chi_{R_{k_1+k_2}}(AB^{\dagger}),\label{LRABJ}
\end{equation}
where the sum is over all Young diagrams in the direct product of $R_1$ and $R_2$, and coefficients $g(R_{k_1},R_{k_2};R_{k_1+k_2})$ give the multiplicity of $R_{k_1+k_2}$ in this product (Littlewood-Richardson coefficients). In fact all of our gauge theory results in this work will follow from \eqref{2ptSchurs} and \eqref{LRABJ}.

\medskip
The program of understanding dynamics of probe branes in M and type IIA theories from the perspective of ABJM was initiated in \cite{Berenstein:2008dc,SheikhJabbari:2009kr,Nishioka:2008ib,Berenstein:2009sa}. However the dictionary involving \eqref{ABJS} started developing only recently \cite{Giovannoni:2011pn,Hamilton:2009iv,Murugan:2011zd,Gutierrez:2010bb,Lozano:2011dd,Herrero:2011bk}. Classical solutions dual to Schurs in $AB^{\dagger}$ are a bit more involved than their counterparts in $AdS_5/CFT_4$. A gravity dual of the symmetric Schur is conveniently described in M-theory as a M2-brane (D2 in IIA) which wraps an $S^2\subset AdS_4$ and rotates along a great circle of $S^7$ orthogonal to the compactification circle. The antisymmetric operator is dual to a D4-brane on $\mathbb{CP}^3$ whose world-volume pinches off as it size increases till it splits into two distinct $D4$s wrapping separate $\mathbb{CP}^2\subset \mathbb{CP}^3$ and intersecting on $\mathbb{CP}^1$. In M-theory this is translated to an M5-brane wrapping two $S^5$'s intersecting at $S^3\subset S^7$ and also rotating on the circle orthogonal to the compactification plane. For more details see \cite{Giovannoni:2011pn,Hamilton:2009iv,Murugan:2011zd,Gutierrez:2010bb,Lozano:2011dd,Herrero:2011bk}.

Giants dual to Schurs in ABJ require taking into account additional NS B field in the type IIA background and up to date no giant gravitons or dual giants have been found with this complication.

\section{ABJ(M) correlators from Schurs }\label{ABJMSchurs}
This section contains our main technical result.  We compute two, three and four-point extremal correlators of single trace half-BPS operators in ABJ(M) in the free field theory limit using Schur polynomial technology. Based on these formulas we write the general form of the n-point correlator. Analogous results in $\mathcal{N}=4$ SYM have been known for a long time \cite{Kristjansen:2002bb,Beisert:2002bb} and played important roles in matching dual observables. 

The n-point extremal correlators of the half-BPS chiral primary operators in ABJ(M) that we compute here are defined as 
\begin{equation}
C^{J_1,...,J_{n-1}}_{n}\equiv\langle \Tr((AB^{\dagger})^{J_1})\ldots \Tr((AB^{\dagger})^{J_{n-1}})\Tr((BA^{\dagger})^{J_n})\rangle,\qquad \sum^{n-1}_{i=1}J_{i}=J_{n}.\label{CorNpt}
\end{equation}

Our main tool will be the formula for the extremal n-point correlation functions in terms of the two point correlator of Schurs labelled by hooks. Such expression can be derived in a general class of gauge theories where the Schur basis can be constructed in some matrix $X$ and the single trace operators are linear combinations of Schurs labeled by hooks\footnote{In ABJ(M) we have \begin{equation}\Tr\left((AB^{\dagger})^{J}\right)=\sum^{J}_{k=1}(-1)^{J-k}\chi_{h^{k}_{J}}(AB^{\dagger})\end{equation}}. The formula reads
\begin{equation}
C^{J_1,...,J_{n-1}}_n=\left(\prod^{n-1}_{l=1}\sum^{J_l}_{k_l=1}\right)\sum^{n-2}_{i=0}(-1)^{i}\left(\begin{array}{c}
n-2 \\
i \end{array}\right)f_{h^{k_{n}-i}_{J_{n}}}\label{NPtI}
\end{equation}
where $J_{n}=\sum^{n-1}_{i=1}J_{i}$, $k_{n}=\sum^{n-1}_{l=1}k_{l}$, and $f_{h^{k}_{J}}$ is the value of the two point correlator of Schur polynomials in $X$ labeled by hooks
\begin{equation}
\langle \chi_{h^{l}_{J}}(X)\chi_{h^{k}_{J}}(\bar{X})\rangle=\delta_{l,k}\,f_{h^{k}_{J}},
\end{equation}
where $h^{k}_{J}$ is the hook of length $J$ with $k$ boxes in the first row. \\
A short outline of the proof of this result is the following. Single trace operators in the extremal n-point correlator can be expressed in a basis of appropriate Schur polynomials and the coefficients of this expansion are non-zero only when diagrams that label Schurs are hooks. The sums over possible Young diagrams become sums over the number of boxes $k_i$ in the first row of each hook, $i=1,...,n$. Moreover, by applying the Littlewood-Richardson fusion rule enough times, the extremal correlator of Schurs can be always written as a linear combination fo the Schur's two point functions. The crucial observation is then that all of the Littlewood-Richardson coefficients must be evaluated on hooks and for given two hooks $h^{k_1}_{J_1}$ and $h^{k_2}_{2}$ coefficient $g(h^{k_1}_{J_1},h^{k_2}_{J_2};h^{l}_{J_1+J_2})$ is non-zero only for $l=\{k_1+k_2,k_1+k_2-1\}$. Finally in every term we get rid of the coefficients by solving for $k_n$ and we arrive at \eqref{NPtI}. More pedagogical details as well as the constructive proof of \eqref{NPtI} in the $\mathcal{N}=4$ SYM context are given in appendix \ref{CorrelatorsN4} .

\medskip
The only input for \eqref{NPtI} is then the two point function of the ABJ Schurs labeled by $h^{k}_{J}$. From \eqref{2ptSchurs} we can easily find that all we need is
\begin{eqnarray}
f_{h^{k}_{J}}(N,M)&=&\prod^{k}_{i=1}(N-1+i)(M-1+i)\prod^{J-k}_{j=1}(N-j)(M-j)\nonumber\\
&=&\frac{\G(N+k)}{\G(N-J+k)}\frac{\G(M+k)}{\G(M-J+k)},\label{fABJ}
\end{eqnarray}
where $\Gamma$ is the Euler Gamma function. It is clear that \eqref{fABJ} is just the product of the two $\mathcal{N}=4$ results \eqref{fN4} for N and M. However, as we will see in following sections, this "squaring" does not carry to the level of the observables (as one could naively expect). The only important message from this structure is that in ABJ we are formally dealing with two Young diagrams that constrain each other. This will become clear in the details of the correlators.

\medskip
Formulas \eqref{NPtI}\footnote{It is actually easier to use slightly more explicit version \eqref{NPtEx}} and \eqref{fABJ} provide the expansion of \eqref{NPtI} for arbitrary $n$ at tree level but to all orders in N and M.  In addition, it is also possible to rewrite these answers in a similar form to \cite{Beisert:2002bb} and below we do that in ABJ for $n=2,3,4$ and provide the general n expression. Results for ABJM model are obtained by setting $N=M$.

\subsection{Two-point functions}
For $n=2$ our formula yields
\begin{equation}
C^{J}_2=\sum^{J}_{k=1}\frac{\G(N+k)\,\G(M+k)}{\G(N-J+k)\,\G(M-J+k)}.
\end{equation}
The sum can be formally performed using Mathematica and the result is
\begin{eqnarray}
C^{J}_2=G(N,M;J)-G(N+J,M+J;J),
\end{eqnarray}
where by $G(a,b;c)$ we denote a special case of the Meijer G-function\footnote{A more general definition and further details on Meijer G-functions can be found in \cite{Gradshteyn}} $G^{1,3}_{3,3}$ that is expressed in terms of the generalized hypergeometric function $_3F_2$ as
\begin{eqnarray}
G(a,b;c)\equiv\frac{\G(a+1)\G(b+1)}{\G(a-c+1)\G(b-c+1)} \,_3F_2\left(\begin{array}{ccc}
1 & a+1 & b+1 \\
\, & a-c+1 & b-c+1  \end{array};1\right).
\end{eqnarray}

\medskip
For large $N$ and $M$ we can expand the two point function and reproduce the known, leading,  result and first sub-leading corrections
\begin{eqnarray}
C^{J}_2\sim J(NM)^{J}\left(1+\frac{J^2(J^2-1)}{12N\,M}+\frac{J(J^2-1)(J-2)}{24N^2}\right.\nonumber\\
\left.+\frac{J(J^2-1)(J-2)}{24M^2}+\ldots\right),\label{2ptABJexp}
\end{eqnarray}
where ellipsis stand for terms of order $J^8N^{-3}M^{-1}$, $J^8N^{-2}M^{-2}$ and $J^8N^{-1}M^{-3}$ etc. 
Notice that the structure is more involved than in $\mathcal{N}=4$ SYM, but again, it clearly shows that if $J$ extends order $\sqrt{N}$ or $\sqrt{M}$, the ranks of the gauge groups are not good expansion parameters anymore.\\
For ABJM the three subleading contributions collapse into one of order $O(N^{-2})$ hence expansion is sensible only for $J$ smaller than $O(\sqrt{N})$.

\subsection{Three-point functions}
Similarly, setting $n=3$ in our formula (or its more explicit form \eqref{NPtEx}) gives
\begin{eqnarray}
C^{J_1,J_2}_3=\left(\sum^{J_1+J_2}_{k=J_2+1}-\sum^{J_1}_{k=1}\right)\frac{\G(N+k)\,\G(M+k)}{\G(N-J_3+k)\,\G(M-J_3+k)},
\end{eqnarray}
where $J_3=J_1+J_2$. Mathematica can formally sum it into a combination of the Meijer G-functions
\begin{eqnarray}
C^{J_1,J_2}_3=G(N+J_1,M+J_1;J_3)+ G(N+J_2,M+J_2;J_3)\nonumber\\
-G(N,M;J_3)-G(N+J_3,M+J_3;J_3)
\end{eqnarray}
To the leading order in $N$ and $M$ and the answer is
\begin{eqnarray}\label{leading-terms-three-point}
C^{J_1,J_2}_3=J_1\,J_2\,J_3\,N^{J_1+J_2-1}M^{J_1+J_2}+\, J_1\,J_2\,J_3\,N^{J_1+J_2}M^{J_1+J_2-1}+...
\end{eqnarray}
Using this result we can compute the normalized three point functions
\begin{eqnarray}
\frac{C^{J_1,J_2}_3}{\sqrt{J_1\,J_2\,J_3}\,(NM)^{J_1+J_2+J_3}}=\frac{\sqrt{J_1\,J_2\,J_3}}{N}+\frac{\sqrt{J_1\,J_2\,J_3}}{M}+...
\end{eqnarray}
This leading contribution was also computed in \cite{Chakrabortty:2011fd} and it is equal to the sum of two leading $\mathcal{N}=4$ three point functions in $N$ and $M$.

\medskip
Three point correlators in ABJM are again obtained by setting $N=M$. The leading answer for the normalized three point functions in ABJM is then twice the $\mathcal{N}=4$ SYM counterpart.

\subsection{Four-point functions}
The tree level four point functions in ABJ are obtained by setting $n=4$ in \eqref{NPtEx} and we have
\begin{equation}
 C^{J_1,J_2,J_3}_{4}=\left(\sum^{J_1}_{k=1}-\sum^{J_1+J_2}_{k=J_2+1}-\sum^{J_1+J_3}_{k=J_3+1}+\sum^{J_4}_{k=J_2+J_3+1}\right)\frac{\G(N+k)\,\G(M+k)}{\G(N-J_4+k)\,\G(M-J_4+k)}.
\end{equation}
Performing the sums brings us to
\begin{eqnarray}
 C^{J_1,J_2,J_3}_{4}=G(M,N;J_4)-G(M+J_4,N+J_4;J_4)\nonumber\\
-G(M+J_1,N+J_1;J_4)-G(M+J_2,N+J_2,J_4)\nonumber\\
-G(M+J_3,N+J_3,J_4)+G(M+J_1+J_2,N+J_1+J_2;J_4)\nonumber\\
+G(M+J_1+J_3,N+J_1+J_3;J_4)+G(M+J_2+J_3,N+J_2+J_3;J_4).
\end{eqnarray}
Expanding the answer to the leading order in $N$ and $M$ yields
\begin{equation}
\frac{C^{J_1,J_2,J_3}_4}{\sqrt{J_1J_2J_3J_4}(NM)^{J_4}}=\sqrt{J_1J_2J_3J_4}\left(\frac{J_4-1}{N^2}+\frac{2J_4}{NM}+\frac{J_4-1}{M^2}\right),
\end{equation}
where $J_4=J_1+J_2+J_3$.
\subsection{N-point correlators}
Clearly the above correlation functions exhibit an interesting structure that can be naturally generalized into and arbitrary n-points. Namely, the tree level $n$-point correlator of the half-BPS chiral primary operators in ABJ can be formally written in terms of Meijer G-functions as
\begin{eqnarray}
C^{J_1,...,J_{n-1}}_n=(-1)^{n}\left[ G(M,N;J_{n})-\sum^{n-1}_{i=1}G(M+J_i,N+J_i;J_n)\right.\nonumber\\
\left.+\sum_{1\leq i_1\leq i_2 \leq n-1}G(M+J_{i_1}+J_{i_2},N+J_{i_1}+J_{i_2};J_{n})-\ldots-G(M+J_{n},N+J_{n};J_{n})\right]\label{NMG}
\end{eqnarray}
where $J_{n}=\sum^{n-1}_{i=1}J_i$, and the ellipsis denote possible terms where arguments of $G$ contain a sum of  three, four, etc. $J$'s with appropriate sign.
Analogous results for ABJM are obtained by setting $N=M$ in the above formula.

A relevant comment is in order at this point. In $\mathcal{N}=4$ SYM n-point extremal correlators of CPOs are protected (see recent proof \cite{Baggio:2012rr}) and the tree level answer is exact. On the contrary, in ABJ(M) they depend on the 't Hooft coupling(s) \cite{Bhattacharya:2008bja}. Determining this coupling dependence is beyond our work but we hope that \eqref{NMG} will serve as a good starting point for understanding the higher loop structure.

In the remaining part of this paper we use these formulas to evaluate various correlators of giants with open and closed strings from the perspective of the ABJ(M) gauge theories. 

\section{Excited giants from ABJ(M)}\label{OpenStrings}
In this section we consider radiation of closed strings from giant gravitons and dual giants in $AdS_4\times \mathbb{CP}^3$, as well as joining and splitting of open strings attached to giants from the gauge theory perspective. This will allow for probing and constraining the giant's geometry. We closely follow \cite{de Mello Koch:2007uu}\footnote{See also \cite{Balasubramanian:2002sa,Balasubramanian:2004nb}} that performed this analysis for $\mathcal{N}=4$ SYM.

\medskip
Based on the experience from $AdS_5/CFT_4$, it is natural to propose ABJ(M) operators dual to excited giants in the type IIA background as Schur polynomials with strings attached 
\begin{equation}
\chi^{(k)}_{R,R_1}(A_1B^{\dagger}_{1},W^{(1)},...,W^{(k)})=\frac{1}{(n-k)!}\sum_{\sigma\in S_{n}}\Tr_{R_1}(\G_R(\sigma))\Tr(\sigma (A_1B^{\dagger}_1)^{\otimes n-k}W^{(1)}...W^{(k)}),
\end{equation}
where $\Tr_{R_1}(\G_R(\sigma))$ is the trace over the subspace of representation $\Gamma_R$, trace with strings is defined as
\begin{equation}
\Tr(\sigma (A_1B^{\dagger}_1)^{\otimes n-k}W^{(1)}...W^{(k)})=(A_1B^{\dagger}_1)^{i_1}_{i_{\sigma(1)}}...(A_1B^{\dagger}_1)^{i_n-k}_{i_{\sigma(n-k)}}(W^{(1)})^{i_n-k+1}_{i_{\sigma(n-k+1)}}...(W^{(1)})^{i_n}_{i_{\sigma(n)}},
\end{equation}
and the strings are represented by words of $A_2B^{\dagger}_2$ of an arbitrary length $J$
\begin{equation}
(W^{(i)})^{i}_{j}=\left((A_2B^{\dagger}_2)^{J}\right)^{i}_{j}.
\end{equation}
Notice that because of the two different ranks $N$ and $M$ we have two different families of strings. In other words, the building blocks of our operators can be arranged in two distinct orders that form matrices of a different size
\begin{equation}
(AB^{\dagger})_{N\times N},\qquad (B^{\dagger}A)_{M\times M},
\end{equation}
we call the first type "N-strings" and the second "M-strings". A consequence of the gauge invariance is that N-strings can be only attached to the $N\times N$ and M-strings to the  $M\times M$ product of $A_1$s and $B^{\dagger}_1$s respectively.

\subsection{Emission of closed strings from giant gravitons}
In this subsection we study the emission of closed strings from the giant gravitons $(\subset \mathbb{CP}^3)$. This is done by evaluating the leading answer to the two point correlator between a Schur with one open string attached and the bound state operator of a Schur and a closed string.

\medskip
The operators dual to excited giant gravitons in ABJ theory are given by Schur polynomials with $O(N)$ rows and $O(1)$ columns with open string attached. Namely, the operator dual to a giant graviton with momentum $p$ with one string of momentum $J$ attached is
\begin{equation}
\chi_{h^{1}_{p+1},\,h^{1}_{p}}^{(1)}(A_{1}B_{1}^{\dagger},\, A_{2}B_{2}^{\dagger})=\frac{1}{(n-1)!}\sum_{\sigma\in S_{n}}\Tr_{h^{1}_{p}}(\G_{h^{1}_{p+1}}(\sigma))\textrm{Tr}((A_{1}B_{1}^{\dagger})^{\otimes n-1}(A_{2}B_{2}^{\dagger})^{J}),
\end{equation}
where the superscript $(1)$ refers to one open string attached.\\
The operator dual to a D-brane (M-brane) with closed string emitted is given by
$$
\textrm{Tr}((A_{2}B_{2}^{\dagger})^{J})\chi_{h^{1}_{p}}(A_{1}B_{1}^{\dagger}).
$$
The amplitude ${\cal{A}}$ that describes the interaction of a D-brane and the giant graviton is thus
\begin{equation}
\mathcal{A}_{h^{1}_{p+1},h^{1}_{p}}=\frac{\bigl\langle\textrm{Tr}((B_{2}A_{2}^{\dagger})^{J})\chi_{h^{1}_{p}}^{\dagger}(B_{1}A_{1}^{\dagger})\chi_{h^{1}_{p+1},\,h^{1}_{p}}^{(1)}(A_{1}B_{1}^{\dagger},\, A_{2}B_{2}^{\dagger})\bigr\rangle}{\bigl\Vert\textrm{Tr}((A_{2}B_{2}^{\dagger})^{J})\bigr\Vert\bigl\Vert\chi_{h^{1}_{p}}\bigr\Vert\bigl\Vert\chi_{h^{1}_{p+1},\,h^{1}_{p}}^{(1)}\bigr\Vert}
\end{equation}
In order to evaluate all the ingredients of this amplitude we can repeat the analysis of \cite{de Mello Koch:2007uu} with \eqref{fABJ} for a single column. Using \eqref{2ptABJexp} and the correlators of open strings that we derived in appendix \ref{Details}, gives the following result for the two point correlator
\begin{equation}
\bigl\langle\textrm{Tr}((B_{2}A_{2}^{\dagger})^{J})\chi_{h^{1}_{p}}^{\dagger}(B_{1}A_{1}^{\dagger})\chi_{h^{1}_{p+1},\,h^{1}_{p}}^{(1)}(A_{1}B_{1}^{\dagger},\, A_{2}B_{2}^{\dagger})\bigr\rangle=J M^{J}N^{J-1}f_{h^{1}_{p+1}}(N,M),
\end{equation}
and the norm of the excited giant
\begin{equation}
\Vert\chi_{h^{1}_{p+1},\,h^{1}_{p}}^{(1)}\bigr\Vert=\left(\frac{p+1}{M}+(J-1)\left(1-\frac{p}{N}\right)\left(1-\frac{p}{M}\right)\right)M^{J+1}N^{J-1}f_{h^{1}_{p+1}}(N,M).
\end{equation}
This norm differs significantly from the counterpart in \cite{de Mello Koch:2007uu} and now the leading contribution comes from the second term\footnote{In \cite{de Mello Koch:2007uu} this $F_0$ contribution was always subleading}.\\
Therefore, the leading contribution to the amplitude for emission of a closed string from the giant graviton is given by
\begin{equation}
\mathcal{A}_{h^{1}_{p+1},h^{1}_{p}}\sim\sqrt{\frac{J(1-\frac{p}{N})(1-\frac{p}{M})}{\frac{p+1}{M}+(J-1)(1-\frac{p}{N})(1-\frac{p}{M})}}\,.
\end{equation}
Similarly to $\mathcal{N}=4$ SYM the amplitude is of order unity for small momenta $p$. However, in ABJ, it only exists for $p$ smaller than $M$. This is a manifestation of the  the bound from the number of boxes in a single column$p\leq M$. Namely, completely antisymmetric Schur polynomials only exist when the number of boxes does not exceed $min(N,M)$. Moreover for maximal giants ($p=M$)  the amplitude vanishes. This can be understood as the consequence of the fact that we cannot excite maximally excited giant by attaching more strings to it.

\medskip
Notice also, that the amplitude is not invariant under the exchange of $M\leftrightarrow N$. This is another manifestation of the parity breaking by subleading corrections that is a known subtlety of the ABJ model (see also \cite{Caputa:2009ug,deMelloKoch:2012kv}).

\subsection{Emission of closed strings from dual giants}
The AdS giant graviton of ABJ theory can be obtained in a similar way. In this case the representation $R$ of ABJ Schur polynomials is the symmetric representation, we denote it by hook $h^{p}_p$. The amplitude is
\begin{align}
\mathcal{A}_{h^{p+1}_{p+1},h^{p}_{p}}\sim\sqrt{\frac{J(1+\frac{p}{N})(1+\frac{p}{M})}{\frac{p+1}{M}+(J-1)(1+\frac{p}{N})(1+\frac{p}{M})}}.
\end{align}
The AdS giant amplitude agrees with the sphere giant amplitude for small $p$ as expected. However it is non zero for the maximal case $p=M$ and slowly decreases for large $p$.

\subsection{String splitting and joining}
To study the splitting and joining of open strings, we need to compute the amplitude of two ABJ Schur polynomials one with string attached and the other with two strings attached. The relevant amplitude describing this process is
\begin{equation}
\mathcal{A}=\frac{\bigl\langle(\chi_{S,S^{\prime}}^{(1)})^{\dagger}\chi_{R,R^{\prime\prime}}^{(2)}\bigr\rangle}{\bigl\Vert\chi_{S,S^{\prime}}^{(1)}\bigr\Vert\bigl\Vert\chi_{R,R^{\prime\prime}}^{(2)}\bigr\Vert},
\end{equation}
where $\chi_{R,R^{\prime\prime}}^{(2)}$ is the ABJ Schur with two strings attached given by
$$
W^{(1)}=(A_2B^{\dagger}_2)^{J_1},\qquad W^{(2)}=(A_2B^{\dagger}_2)^{J_2}.
$$
$\chi_{S,S^{\prime}}^{(1)}$ has one string attached and it is given by
$$
W=(A_2B^{\dagger}_2)^{J_1+J_2}.
$$
Labels $R$ and $S$ denote Young diagrams labeling Schurs dual to giants and $S'$ and $R''$ stand for diagram $S$ with one box removed and diagram $R$ with two boxes removed.\\
The amplitude can be computed using appendix H of \cite{de Mello Koch:2007uu} for ABJ operators together with the results obtained in our appendix D, we consider the case where both $R$ and $S$ have rows of $O(N)$ and columns of $O(1)$ (the other case follows directly by changing to the symmetric representation). We consider the first column of $R$ has a length $b_1+b_2$ where $b_1$ is the length of the second column and it is $O(N)$, $b_2$ is $O(1)$ (Fig. 1).

\medskip
 \begin{figure}[!h]
       \begin{center}
     \includegraphics[scale=1]{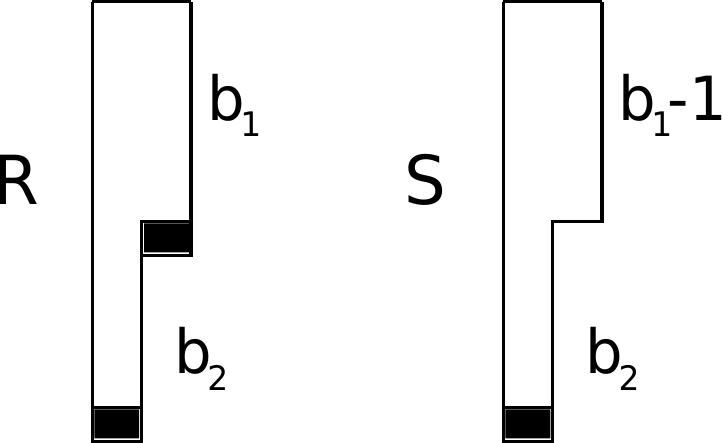} 
     \caption{\textit{Young diagrams used in the computation of the string joining amplitude}}
     \end{center}
     \label{YTab}
     \end{figure}

 The leading contribution to the numerator comes from the terms that contain $C_5$ and $C_6$ in appendix D. The result is
\begin{equation}
\mathcal{A}=2\sqrt{\frac{(N-b_{1})(M-b_{1})}{b_{1}NM}}\frac{1}{(b_{2}+1)}\left(1+O(J^{8}(NM)^{-2})\right).
\end{equation}
We note that this amplitude is independent of the angular momentum of the open strings. Moreover, this leading answer is invariant under the exchange of $M\leftrightarrow N$.

\subsection{Predictions for dual probes}
Analyzing the above amplitudes one can propose the following qualitative picture for the dual probe branes on $\mathbb{CP}^3$ with B2 field. Naturally, the geometry has to generalize the dual ABJM giant graviton found in \cite{Giovannoni:2011pn} that is described by two $D4$-branes wrapping separate $\mathbb{CP}^2$s. In ABJ the two different ranks should correspond to two different radius sizes of the $\mathbb{CP}^2$s. If we then want to attach strings into the $D4$s, N-strings end on the space with radius N and M-strings on the space with radius M (see Figure 2.). Moreover, from the gauge no open strings can stretch between $D4$-branes on the two different spaces. This constraint disappears when we set $M=N$ for ABJM. 

\medskip
 \begin{figure}[!h]
       \begin{center}
     \includegraphics[scale=0.6]{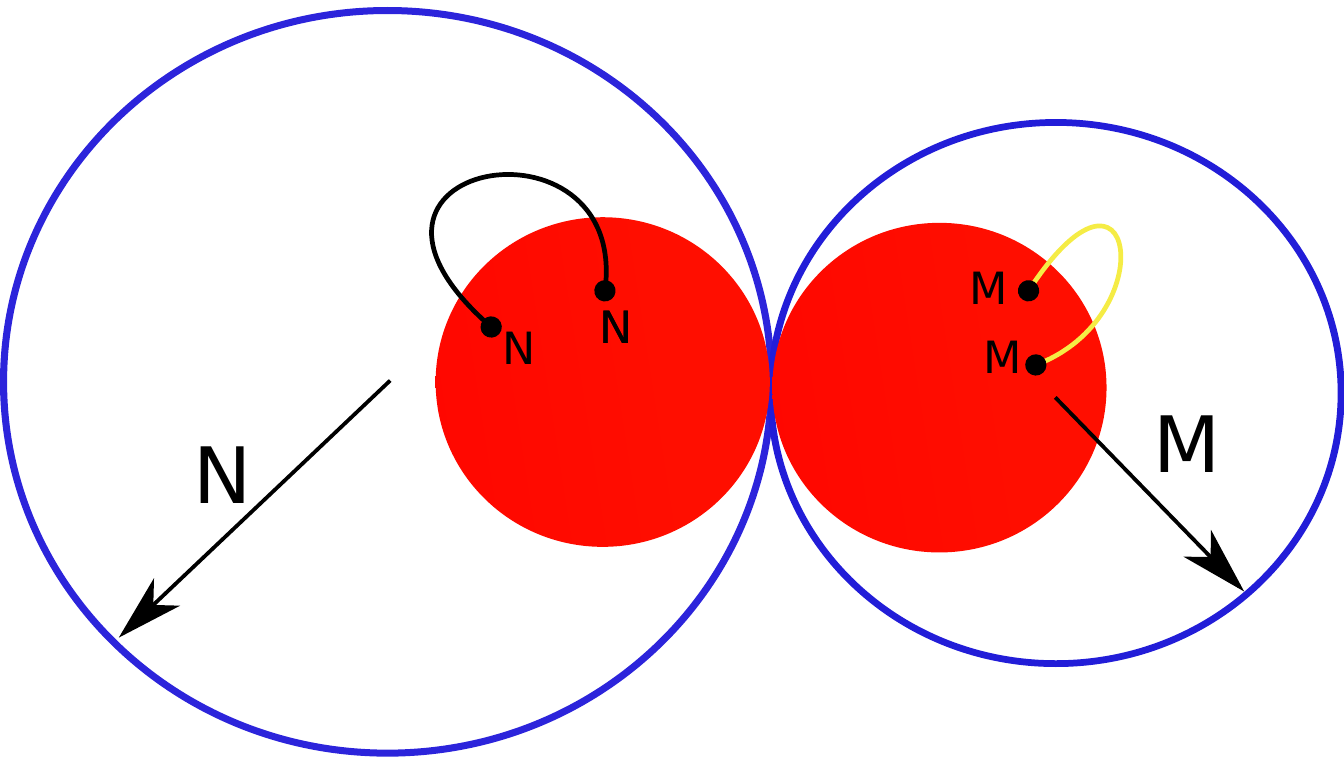} 
     \caption{\textit{Possible geometrical dual of the antisymmetric ABJ Schurs. Two different ranks in the ABJ gauge group might be interpreted as two different radii of the two $\mathbb{CP}^2$s. A cut off on the number of boxes of the Young diagrams labeling the gauge theory operator $(k\leq M)$ is then naturally realized in the dual geometry. Gauge invariance requires that end points of strings must be attached to the part in space with radius N (N-strings) or M (M-strings), and no strings can stretch between the two separate parts. }}
     \end{center}
     \label{GGABJ}
     \end{figure}

\section{More correlators in ABJ(M)}\label{MoreCorrelators}

In order to provide more data into the dictionary between schurs and giants in ABJ(M), in this section, we compute several additional correlation functions with Schur polynomials and CPOs in ABJ. Our results are valid for arbitrary possible Young diagrams labelling Schur polynomials but we also consider special cases where diagrams have the form of a single column or a single row. Moreover, we discuss a limit where our results give predictions for semiclassical giant gravitons or dual giants in the $AdS_4\times \mathbb{CP}^{3}$ geometry with additional background NS B-field. For $N=M$ we reduce to the ABJM theory.

\subsection{Schur and a CPO}
The simplest correlator that we start with is the two point function of one Schur and a CPO
\begin{equation}
C^{R_J}_{12}=\frac{\langle \chi_{R_J}(AB^{\dagger})\Tr\left((A^{\dagger}B)^{J}\right)\rangle}{||\chi_{R_J}(AB^{\dagger})||\,||\Tr\left((A^{\dagger}B)^{J}\right)||},
\end{equation}
where $R_J$ is an arbitrary Young tableaux with $J$ boxes. For completely symmetric or antisymmetric representations it describes a transition\footnote{For discussion and interpretation of "transition" processes involving half-BPS operators see \cite{Brown:2006zk}} between point-like graviton and the giant graviton or the dual giant. 

\medskip
We proceed with the computation in the usual way by expanding the single trace operator in a Schur basis with characters evaluated on a J-cycle permutations. Inserting the explicit formulas for the norms and the two point correlator \eqref{2ptSchurs} yields
\begin{equation}
C^{R_J}_{12}=\sum_{T_J}\chi_{T_J}(\sigma_{J}) \delta_{R_JT_J}\frac{f_{T_{J}}(N,M)}{\sqrt{f_{R_{J}}(N,M)\,J(NM)^J}}.
\end{equation}
Notice that now delta function projects on the term $T_J=R_J$ but also the character of the J-cycle implies that we only get a non-vanishing contribution when $R_J$ is a hook with $J$ boxes \cite{Fulton}. We parametrize hooks by the number of boxes $k$ in the first row and denote it by $R_J=h^{k}_J$. The final answer is
\begin{equation}
C^{h^{k}_J}_{12}=\frac{(-1)^{J-k}}{\sqrt{J}}\sqrt{\frac{f_{h^{k}_{J}}(N,M)}{(NM)^J}},
\end{equation}
where $ f_{h^{k}_{J}}$ is the product over weights of the hook \eqref{fABJ}. 

\medskip
The lesson is, that the transition to a point-like graviton is only possible for objects that are dual to Schurs labelled by hook diagrams \footnote{Note that our computation would be completely analogous for $\mathcal{N}=4$ SYM hence the same conclusion holds in $AdS_5/CFT_4$.}. We can explicitly analyze this process for the M2 and the M5 branes (or D2 and D4) and the point-like graviton moving on the circle orthogonal to the compactification plane. This are of course the two special cases of hooks, the completely symmetric $k=J$ and antisymmetric $k=1$ representations the answers become
\begin{eqnarray}
C^{h^{J}_J}_{12}&=&\sqrt{\frac{1}{J}+\frac{J}{2}\left(\frac{1}{N}+\frac{1}{M}\right)},\nonumber\\
 C^{h^{1}_J}_{12}&=& (-1)^{J-1}\sqrt{\frac{1}{J}-\frac{J}{2}\left(\frac{1}{N}+\frac{1}{M}\right)}.
\end{eqnarray}
For $J\sim O(1)$ both transitions are suppressed as $1/\sqrt{J}$. This is in agreement with $AdS_5/CFT_4$ and to the leading order the transition is independent on the fact that a brane wraps the cycle inside $AdS_4$ or $\mathbb{CP}^3$.  Again the amplitude with Schurs in the antisymmetric representation is only valid for $J$ smaller than $M$. This is another manifestation of the bound on the giant gravitons in $\mathbb{CP}^3$. If $J\sim O(M)$ for the dual giant, probability for this transition is of order $O(1)$.

\subsection{Schur and two CPOs}
Another interesting correlators are the three point functions with one Schur polynomial and two CPOs
\begin{equation}
C^{R_k}_{123}=\frac{\langle \chi_{R_k}(AB^{\dagger})\Tr\left((BA^{\dagger})^{k-J}\right)\Tr\left((BA^{\dagger})^{J}\right)\rangle}{||\chi_{R_J}(AB^{\dagger})||\,||\Tr\left((BA^{\dagger})^{k-J}\right)||\,||\Tr\left((BA^{\dagger})^{J}\right)||}.
\end{equation}
For large $k$ and $J$ they describe the decay of brane objects into semiclassical closed strings (gravitons) and for $J$ of order $O(1)$ one of the strings becomes very light so we have a semiclassical transition of a giant into a point-like graviton probed by a light (quantum) supergravity mode.

\medskip
Again we first expand the CPOs in a Schur basis, then use the Littlewood-Richardson rule for the product of two Schurs and finally use fact that the characters on J-cycles are non-zero only for hooks which we denote $h^{p}_{k-J}$ and $\tilde{h}^{q}_{J}$. We then have a general result valid for any $R_{k}$ that appears in a direct product of two hooks
\begin{eqnarray}
C^{R_k}_{123}=\sqrt{\frac{f_{R_k}(N,M)}{J(k-J)N^kM^k}}\sum^{k-J}_{p=1}\sum^{J}_{q=1}(-1)^{k-p-q}\,g(h^{p}_{k-J},\tilde{h}^{q}_{J};R_k)
\end{eqnarray}
where $g(h^{p}_{k-J},\tilde{h}^{q}_{J};R_k)$ is the Littlewood-Richardson coefficient that gives the multiplicity of diagram $R_k$ in a direct product of hooks $h^{p}_{k-J}$ and $h^{q}_{J}$.

For completely symmetric and antisymmetric representations we have
\begin{eqnarray}
C^{h^{k}_k}_{123}&=&\sqrt{\frac{1}{J(k-J)}+\frac{k^2}{2J(k-J)}\left(\frac{1}{N}+\frac{1}{M}\right)}
\nonumber\\
 C^{h^{1}_k}_{123}&=&(-1)^{k}\sqrt{\frac{1}{J(k-J)}-\frac{k^2}{2J(k-J)}\left(\frac{1}{N}+\frac{1}{M}\right)}
\end{eqnarray}
This clearly shows that when one of the operators is very light we just probe the transition from the previous subsection with some light mode. When both charges, $k$ and $J$ are light we have a similar story to the transition between giants and point-like gravitons and the process is suppressed as inverse of the square root of the charges of the states that giants decay into.  For giants dual to Schurs labeled by a single column the amplitude is only defined for charges not greater than $M$.
\subsection{Two Schurs and a CPO}
Another interesting observables that probe the physics of probe branes form the gauge theory are the three point functions of two Schur polynomials and one chiral primary operator
\begin{equation}
C^{R}_{123}=\frac{\langle \chi_{R_{k}}(AB^{\dagger})\chi_{R_{k-J}}(BA^{\dagger})\,\Tr\left((BA^{\dagger})^{J}\right)\rangle}{||\chi_{R_{k}}(AB^{\dagger})||\,||\chi_{R_{k-J}}(BA^{\dagger})||\,||\Tr\left((BA^{\dagger})^{J}\right)||},
\end{equation}
where $R_{n}$ is either a completely symmetric (row) or a completely antisymmetric (column) Young tableaux with $n$ boxes. Physically, for $J\sim O(1)$, the amplitude describes the emission/absorption of the light mode from/to probe branes neglecting backreaction.\\
Using the results derived in previous sections we have
\begin{equation}
C^{R}_{123}=\chi_{R_J}(\sigma_J)\sqrt{\frac{f_{R_{k}}(N,M)}{f_{R_{k-J}}(N,M)\,J(NM)^{J}}}.
\end{equation}
Once we extract the completely symmetric and antisymmetric answers from \eqref{fABJ} and take the semiclassical limit  
\begin{equation}
N,M\to\infty,\qquad k\to\infty,\qquad \frac{k}{N},\,\frac{k}{M}\quad\text{finite},\qquad J\ll k,
\end{equation}
the coefficients of the three point functions become
\begin{eqnarray}
C^{h^{k}_k}_{123}&=&\frac{1}{\sqrt{J}}\left(1+\frac{k}{N}\right)^{J/2}\left(1+\frac{k}{M}\right)^{J/2},\nonumber\\
C^{h^{1}_k}_{123}&=&(-1)^{J-1}\frac{1}{\sqrt{J}}\left(1-\frac{k}{N}\right)^{J/2}\left(1-\frac{k}{M}\right)^{J/2}.\label{HHL}
\end{eqnarray}
Note that the $N$ ($M$) dependence looks like the two copies of the $\mathcal{N}=4$ SYM structure constants \cite{Bissi:2011dc} but the $J$ factor in front is the same. Recall that on the gravity side, semiclassical three point functions are computed by fluctuating\footnote{With the appropriate supergravity mode dual to the light $J\sim O(1)$ operator} the DBI action evaluated on Wick rotated classical giant graviton solution written in an appropriate coordinates \cite{Janik:2010gc}. In these coordinates, the $\mathcal{N}=4$ giant gravitons are simply classical (tube-like) propagators between two points at the boundary of $AdS_5$. The form of \eqref{HHL} suggests that analog transformation should map giants on $\mathbb{CP}^3$ into "doubled" classical propagators between points at the boundary of $AdS_4$. It would be interesting to verify this doubling in detail once giants dual to Schurs in ABJ are constructed.

If we set $N=M$ our coefficients reduce to the extremal three correlators in ABJM that were studied in \cite{Hirano:2012vz} both in gauge theory and gravity.

\section{Future directions}
We finish by outlining several possible extensions of our work.

\medskip
The tree-level n-point correlation functions of CPOs that we derived in this work are expressed in terms of a particular type of Meijer G-function. It would be interesting to reproduce our result from a matrix model computation in the scalar subsector of ABJ(M). The analog of Ginibre's \cite{Ginibre:1965zz} method that was primarily used in \cite{Kristjansen:2002bb} has been recently worked out for products of matrices in \cite{Akemann}, and can possibly be adjusted to the ABJ(M) context.

Our general formula for the correlators in terms of the two-point function of Schurs labeled by hooks is just another manifestation of how powerful and elegant Schur polynomials are. It would be interesting to investigate how this result modifies in $\mathcal{N}=4$ SYM with $SU(N)$ \cite{de Mello Koch:2004ws} or other gauge groups \cite{Caputa:2010ep,CaputaSON}. Moreover, there has been a lot of progress in extending Schur techniques to multi-matrix polynomials. A basis and the analog of the Littlewood-Richardson rule has been constructed in \cite{Bhattacharyya:2008rb}. Generalization of our formula to this setup would be also very useful especially from the perspective of the recent developments in three-point-functions \cite{Foda:2011rr}.
 
Dictionary between Schur polynomials and probe branes is still in the very early stage of developing. The new $AdS_4/CFT_3$ dualities with their variety of branes and non-trivial geometry serve as an excellent testing ground for understanding of how much information is encoded in Schur operators. The next natural step is to explicitly construct the appropriate giant gravitons in $AdS_4\times\mathbb{CP}^3$ with NS B-field (see \cite{Prokushkin:2004pv} for related construction) and verify how accurate are the predictions from the gauge theory processes studied in this work.

\subsubsection*{Acknowledgments}
We thank Robert de Mello Koch for many enjoyable discussions, suggestions during this project and Agnese Bissi and Robert de Mello Koch for comments on the draft. PC would like to thank Charlotte Kristjansen for support and hospitality at the Niels Bohr Institute where part of this work was performed. This work is based upon research supported by the South African Research Chairs Initiative of the Department of Science and Technology and National Research Foundation.


\begin{appendix}

\section{Extremal correlators from Schurs }\label{CorrelatorsN4}
In this Appendix we review and provide a detailed derivation of extremal two, three, and four point correlators of chiral primary operators (CPO) in $\mathcal{N}=4$ SYM with $U(N)$ gauge group using Schurs. They were first obtained using both, Schur polynomial technology (two and three point) and the matrix model in \cite{Kristjansen:2002bb} (see also \cite{Corley:2002mj},\cite{Corley:2001zk}). A general expression for the n-point correlator was conjectured in \cite{Beisert:2002bb,Okuyama:2002zn}. Here we give a constructive proof of these formulas and express them in terms of weights of hook diagrams. This makes it easily extendable to a larger class of gauge theories (including ABJ(M) models) in which single trace chiral primary operators can be expressed in a basis of Schur polynomials.

Recall that in $\mathcal{N}=4$ SYM chiral primary operators are single-trace symmetrized products of the six scalar fields
\begin{equation}
O^{CPO}_{I}=\frac{1}{\sqrt{JN^{J}}}\,C^{i_{1}\ldots i_{J}}_{I}\,\Tr(\phi_{i_1}\ldots \phi_{i_J}),
\end{equation}
where $C^{i_{1}\ldots i_{J}}_{I}$ are symmetric traceless tensors of $SO(6)$. 

A special class of these operators, so-called the BMN-type chiral primaries,  are the highest-weight states in the $[0,J,0]$ representation of $SO(6)$ and are expressed in terms of one complex scalar
\begin{equation}
O_{J}=\frac{1}{\sqrt{JN^{J}}}\,\Tr(Z^J),\qquad Z=\phi_{1}+i\,\phi_{2}.\label{CPOZJ}
\end{equation}
These are the operators that we will be concerned with in this part, and in particular their n-point extremal correlators 
\begin{equation}
C^{J_1,...,J_{n-1}}_{n}\equiv\langle \Tr(Z^{J_1})\ldots \Tr(Z^{J_{n-1}})\Tr(\bar{Z}^{J_n})\rangle,\qquad \sum^{n-1}_{i=1}J_{i}=J_{n}.\label{NN4}
\end{equation}
For convenience we drop the normalization factors that can be easily recovered at any stage.

\medskip
Below we evaluate these correlators for general n using the technology of Schur polynomials and here are three sufficient tools to for this task:
\begin{itemize}
\item CPOs can be expanded in a basis of Schur polynomials\footnote{see \cite{Corley:2002mj},\cite{Corley:2001zk} for more details on the Schur polynomial basis}
\begin{equation}
\Tr(Z^{J})=\sum_{R}\chi_{R}(\sigma_{J})\,\chi_{R}(Z),
\end{equation}
where the sum is over all possible Young diagrams with $J$ boxes, $\chi_{R}(\sigma_{J})$ is a character of the $J$-cycle permutation in representation $R$ and $\chi_{R}(Z)$ is the Schur polynomial in matrix $Z$ that transforms in the adjoint representation of $U(N)$ (for $SU(N)$ see \cite{Corley:2002mj,de Mello Koch:2004ws}). It is also a known fact that characters of the $J$-cycle permutations are non-vanishing only for hook diagrams \cite{Fulton}, and for a hook with $k$ boxes in the first row we have
\begin{equation}
\chi_{h^{k}_{J}}(\sigma_{J})=(-1)^{J-k},\label{charh}
\end{equation}
where we denote the hook of length $J$ with $k$ boxes in the first row by $h^{k}_{J}$. This way the sum over $R$ can be written as sum over the $k$, and we have
\begin{equation}
\Tr(Z^{J})=\sum_{k}(-1)^{J-k}\,\chi_{h^{k}_{J}}(Z).\label{expansionS}
\end{equation}
\item The two point correlator of Schurs is given by\cite{Corley:2002mj}
\begin{equation}
\langle \chi_{R_1}(Z)\chi_{R_2}(\bar{Z})\rangle=\delta_{R_1 R_2}\,f_{R_2}(N),\label{ortN4}
\end{equation}
where $f_{R}(N)$ is the product over the weights of the Young diagram.
\item The Littlewood-Richardson rule \cite{Fulton} states that the product of two Schur polynomials with $J_1$ and $J_2$ boxes can be expressed as a linear combination of Schurs with $J_1+J_2$ boxes
\begin{equation}
\chi_{R_1}(Z)\chi_{R_2}(Z)=\sum_{T_3}g(R_1,R_2;T_3)\chi_{T_3}(Z),
\end{equation}
where the Littlewood-Richardson coefficients $g(R_1,R_2;T_3)$ give the multiplicity of the representation $T_3$ in the tensor product of representations $R_1$ and $R_2$.
\end{itemize}

\medskip
 The constructive proof of the general form of \eqref{NN4} can then be obtained by applying the following \text{\bf Algorithm}:
\begin{itemize}
\item Start by expressing all the CPOs in the basis of Schurs using \eqref{expansionS}

\item Next, use the Littlewood-Richardson enough times that the answer is expressed as a linear combination of the results of the two point function of Schurs labeled by hooks with $J_n$ boxes. Coefficients of this linear combination will be $g(R_1,R_2;T_3)$ with all entries given by hooks (this is valid only for extremal correlators).

\item Finally, somewhat simple and easy to check (Appendix \ref{Hooks}) but a crucial observation is that there are only two possible hooks that can appear in a tensor product of two hook diagrams (other diagrams are not hooks). Namely for hooks with $k_1$ and $k_2$ boxes the two hooks in the direct product have $k_1+k_2$ and $k_1+k_2-1$ boxes in the first row. 
\item Use this fact to get rid of the Littlewood coefficients obtaining the elegant answer only in terms of $f_{h^k_{J_{n}}}$.
\end{itemize}

Below we demonstrate how the algorithm works in practise and how it yields the n-point correlators conjectured in \cite{Beisert:2002bb,Okuyama:2002zn}.

\medskip
Let us start with the two point correlator. We first express the two point function of the CPO's in a basis of Schurs using \eqref{expansionS}
\begin{equation}
C^{J}_2=\langle \Tr\left(Z^{J}\right)\Tr\left(\bar{Z}^{J}\right)\rangle=\sum^{J}_{k_{1},k_{2}=1}(-1)^{2J-k_1-k_2}\langle \chi_{h^{k_1}_{J}}(Z)\chi_{h^{k_2}_{J}}(\bar{Z})\rangle.
\end{equation}
 For hooks $\delta_{R_1,R_2}$ becomes $\delta_{k_1,k_2}$ and we have
\begin{equation}
C^J_2=\sum^{J}_{k=1}f_{h^{k}_{J}}(N)\label{2ptN4sum},
\end{equation}
hence the correlator is just the sum over weights of all the possible hook diagrams with $J$ boxes. In $\mathcal{N}=4$ SYM with $U(N)$ gauge group, the product over weights of the hook is given by 
\begin{equation}
f_{h^{k}_{J}}(N)=\prod^{k}_{i=1}(N-1+i)\prod^{J-k}_{m=1}(N-m)=\frac{\G(N+k)}{\G(N-J+k)},\label{fN4}
\end{equation}
where $\Gamma$ is the Euler Gamma function. Inserting this to \eqref{2ptN4sum} reproduces the two point function of the BMN-type CPO's \cite{Kristjansen:2002bb}
\begin{eqnarray}
C^J_2=\frac{1}{J+1}\left(\frac{\Gamma(N+J+1)}{\Gamma(N)}-\frac{\Gamma(N+1)}{\Gamma(N-J)}\right).
\end{eqnarray}
This baby example and the result can be obtained alternatively using Ginibre's method \cite{Ginibre:1965zz}.

Similarly we can proceed with extremal three point functions. First we express the CPOs in terms of Schurs
\begin{equation}
C^{J_1,J_2}_3=\sum^{J_1,J_2,J_3}_{k_1,k_2,k_3=1}\prod^{3}_{i=1}(-1)^{J_i-k_i}\langle\chi_{h^{k_1}_{J_1}}(Z)\chi_{h^{k_2}_{J_2}}(Z)\chi_{h^{k_3}_{J_3}}(\bar{Z})\rangle,
\end{equation}
where $J_3=J_1+J_2$. Next, using \eqref{ortN4} brings us to the sum
\begin{equation}
C^{J_1,J_2}_3=\sum^{J_1,J_2,J_3}_{k_1,k_2,k_3=1}\prod^{3}_{i=1}(-1)^{J_i-k_i}g(h^{k_1}_{J_1},h^{k_2}_{J_2};h^{k_3}_{J_3})f_{h^{k_3}_{J_3}},
\end{equation}
where the crucial ingredient is the Littlewood-Richardson coefficient for hooks. We "kill" it by replacing the sum over $k_3$ by the only two possible terms, $k_3=k_1+k_2$ and $k_3=k_1+k_2-1$. This brings us to
\begin{equation}
C^{J_1,J_2}_3=\sum^{J_1}_{k_1=1}\sum^{J_2}_{k_2=1} \left(f_{h^{k_1+k_2}_{J_3}}-f_{h^{k_1+k_2-1}_{J_3}}\right)=\left(\sum^{J_1+J_2}_{k=J_2+1}-\sum^{J_1}_{k=1}\right)f_{h^{k}_{J_3}},
\end{equation}
where in the second equality we took into account the mutual cancellations between the terms.
Finally, plugging \eqref{fN4} gives the exact three point function of CPOs in $\mathcal{N}=4$ SYM \cite{Kristjansen:2002bb}
\begin{eqnarray}
C^{J_1,J_2}_3=\frac{1}{J_1+J_2+1}\left(\frac{\Gamma(N+J_1+J_2+1)}{\Gamma(N)}+\frac{\Gamma(N+1)}{\Gamma(N-J_1-J_2)}\right.\nonumber\\
\left.-\frac{\Gamma(N+J_1+1)}{\Gamma(N-J_2)}-\frac{\Gamma(N+J_2+1)}{\Gamma(N-J_1)}\right).
\end{eqnarray}

Following the algorithm for four points, we expand CPOs in Schurs and apply the Littlewood-Richardson rule twice what yields\footnote{Note that we use $(-1)^{\sum^{n-1}_{i=1}2J_i-\sum^{n}_{i=1}k_1}=(-1)^{\sum^{n}_{i=1}k_i}$}
\begin{equation}
 C^{J_1,J_2,J_3}_{4}=\left(\prod^{4}_{i=1}\sum^{J_i}_{k_i=1}(-1)^{k_i}\right)\sum_{T}g(h^{k_1}_{J_1},h^{k_2}_{J2};T)\,g(T,h^{k_3}_{J_3};h^{k_4}_{J_4})\,f_{h^{k_4}_{J_4}}.
\end{equation}
By carefully analyzing the Littlewood-Richardson coefficients we can see that there are only two possibilities for $T$ in the first coefficient, namely: $T\in\{h^{k_1+k_2}_{J_1+J_2},h^{k_1+k_2-1}_{J_1+J_2}\}$. These two cases inserted to the second coefficient reduce the sum over $k_4$ into four terms. The first one with $k_4=k_1+k_2+k_3$ with a plus sign, then we have twice $k_4=k_1+k_2+k_3-1$ with a minus sign and $k_4=k_1+k_2+k_3-2$ with plus again. This way the four point correlator becomes
\begin{equation}
 C^{J_1,J_2,J_3}_{4}=\sum^{J_1}_{k_1=1}\sum^{J_2}_{k_2=1}\sum^{J_3}_{k_3=1}\left(f_{h^{k_1+k_2+k_3}_{J_4}}-2f_{h^{k_1+k_2+k_3-1}_{J_4}}+f_{h^{k_1+k_2+k_3-2}_{J_4}}\right).\label{4ptN4Sums}
\end{equation}
Clearly, most of the terms in this sums mutually cancel and it is easy to check that then only ones left can be written as
\begin{equation}
 C^{J_1,J_2,J_3}_{4}=\left(\sum^{J_1}_{k=1}-\sum^{J_1+J_2}_{k=J_2+1}-\sum^{J_1+J_3}_{k=J_3+1}+\sum^{J_4}_{k=J_2+J_3+1}\right)f_{h^{k}_{J_4}}
\end{equation}
Inserting \eqref{fN4}, we can easily perform the sums in Mathematica and a tree level answer for the four point correlator of BMN-type chiral primaries in $\mathcal{N}=4$ SYM is
\begin{eqnarray}
 C^{J_1,J_2,J_3}_{4}=\frac{1}{J_4+1}\left\{\frac{\G(N+J_4+1)}{\G(N)}-\frac{\G(N+J_4-J_1+1)}{\G(N-J_1)}\right.\nonumber\\
\left.-\frac{\G(N+J_4-J_2)}{\G(N-J_2)}-\frac{\G(N+J_4-J_3)}{\G(N-J_3)}+\frac{\G(N+J_4-J_1-J_2)}{\G(N-J_1-J_2)}\right.\nonumber\\
\left.+\frac{\G(N+J_4-J_2-J_3)}{\G(N-J_2-J_3)}+\frac{\G(N+J_4-J_1-J_3)}{\G(N-J_1-J_3)}-\frac{\G(N+1)}{\G(N-J_4)}\right\}.
\end{eqnarray}

It is straightforward to write down an expression for the n-point extremal correlator of half-BPS, single trace CPOs. In fact we can do it for a general class of gauge theories where a basis of Schur polynomials in some unitary matrix $X$ can be constructed and CPOs written as linear combinations of Schurs labeled by hooks. Namely, following our algorithm the crucial step becomes the evaluation of the correlator of Schurs 
\begin{eqnarray}
\langle\prod^{n-1}_{i=1}\chi_{h^{k_i}_{J_i}}(X)\,\chi_{h^{k_n}_{J_n}}(\bar{X})\rangle&=&
\left(\prod^{n-3}_{i=1}\sum_{T_i}\right)g(h^{k_1}_{J_1},h^{k_2}_{J_2};T_1)\left(\prod^{n-4}_{j=1} g(T_j,h^{k_{j+2}}_{J_{j+2}};T_{j+1})\right)\nonumber\\
&\times& g(T_{n-3},h^{k_{n-1}}_{J_{n-1}};h^{k_{n}}_{J_{n}})\,f_{h^{k_n}_{J_n}}.
\end{eqnarray}
Then by taking into account the fact that each sum over $T_i$ contains only two possible hooks and finally performing the sum over $k_{n}$ we end up with our master formula
\begin{equation}
C^{J_1,...,J_{n-1}}_n=\left(\prod^{n-1}_{l=1}\sum^{J_l}_{k_l=1}\right)\sum^{n-2}_{i=0}(-1)^{i}\left(\begin{array}{c}
n-2 \\
i \end{array}\right)f_{h^{k_{n}-i}_{J_{n}}}\label{NPt}
\end{equation}
where $J_{n}=\sum^{n-1}_{i=1}J_{i}$, $k_{n}=\sum^{n-1}_{l=1}k_{l}$, and $f_{h^{k}_{J}}$ is the value of the two point correlator of Schur polynomials in $X$ labeled by hooks of length $J$ with $k$ boxes in the first row
\begin{equation}
\langle \chi_{h^{l}_{J}}(X)\chi_{h^{k}_{J}}(\bar{X})\rangle=\delta_{l,k}\,f_{h^{k}_{J}}.
\end{equation}

If we take into account the cancellations between the terms in \eqref{NPt}, the answer can be written as
\begin{eqnarray}
C^{J_1,...,J_{n-1}}_n=(-1)^{n}\left(\sum^{J_1}_{k=1}-\sum^{J_1+J_2}_{k=J_2+1}-\ldots-\sum^{J_1+J_{n-1}}_{k=J_{n-1}+1}+\sum^{J_1+J_2+J_3}_{k=J_2+J_3+1}+\ldots\right.\nonumber\\
\left.+\sum^{J_{n-2}+J_{n-1}+J_{1}}_{k=J_{n-2}+J_{n-1}+1}-\sum^{J_1+J_{2}+J_{3}+J_{4}}_{k=J_{2}+J_{3}+J_{4}+1}-...+\sum^{J_n}_{k=J_{2}+..+J_{n-1}}\right)f_{h^{k}_{J_n}}.\label{NPtEx}
\end{eqnarray}
This formula can be easily proved by induction\footnote{We thank Robert de Mello Koch for discussion on this point and outlining the proof to us.}. 

For $\mathcal{N}=4$ SYM our formula precisely gives the n-point correlators conjectured in \cite{Beisert:2002bb,Okuyama:2002zn} which are
\begin{eqnarray}
C^{J_1,...,J_{n-1}}_n=\frac{1}{J_n+1}\left\{\frac{\G(N+J_{n}+1)}{\G(N)}-\sum^{n-1}_{i=1}\frac{\G(N+J_{n}-J_i+1)}{\G(N-J_i)}\right.\nonumber\\
\left.+\sum_{1\leq i_1\leq i_2\leq n-1}\frac{\G(N+J_{n}-J_{i_1}-J_{i_2}+1)}{\G(N-J_{i_1}-J_{i_2})}-\ldots-\frac{\G(N+1)}{\G(N-J_{n})}\right\}
\end{eqnarray}
where $J_{n}=\sum^{n-1}_{i=1}J_i$, and ellipsis stand for terms where we subtract more available $J$'s inside $\G$ in the numerator with appropriate sign for even and odd numbers.

\section{Littlewood-Richardson coefficient for hooks} \label{Hooks}
Littlewood-Richardson coefficients $g(R_1,R_2;R_3)$ give the multiplicity of the $U(N)$ representation $R_3$ in the tensor product of representations $R_1$ and $R_2$. When computing higher point correlators of single trace operators with Schurs we only encounter Littlewood-Richardson coefficients with all three $R_i$ in a shape of a hook diagram. In particular, an important result that we needed in order to evaluate the three point correlator of single trace CPOs was that for given two hooks $R_1$ and $R_2$, there are only two hooks in $R_3$, both with multiplicity $g=1$. To demonstrate this, let us first recall the decomposition of a direct product of Young diagrams.

When decomposing a tensor product of two Young diagrams we must obey the following rules \cite{Cvitanovic} : First draw the diagrams next to each other and label boxes of the second diagram with the number of the row to which they belong, counting from the top. Then add labelled boxes to the first (unlabeled) diagram to create new diagrams such that the following holds
\begin{itemize}
\item Each new diagram must be a Young diagram
\item For $U(N)$ no diagram can have more than $N$ rows
\item Passing through the diagram starting from the top row and entering each row from the right, at any point the number of $i$-s  encountered in any of the attached boxes must not exceed the number of previously encountered $(i-1)$-s
\item Numbers must not increase when reading across a row from left to right
\item Numbers must decrease when reading a column from bottom to the top.
\end{itemize}

Following these rules we can see that in a decomposition of a product of two hook diagrams with $l_1$ and $l_2$ boxes in the first rows respectively, we can only have two hooks, one with $l_1+l_2$ and the other with $l_1+l_2-1$ boxes in the first row. See for example
$$
\young({\,}{\,}{\,},{\,},{\,},{\,})\otimes\young({1}{1}{1}{1},{2},{3},{4})
=\young({\,}{\,}{\,}{1}{1}{1}{1},{\,},{\,},{\,},{2},{3},{4})\oplus\young({\,}{\,}{\,}{1}{1}{1},{\,},{\,},{\,},{1},{2},{3},{4})\oplus\ldots
$$

\section{The proof of orthogonality with projectors} \label{Proof}
Proving various properties of Schur polynomials can quickly lead to cumbersome and lengthy formulas. Fortunately there exists a convenient notation in terms of projection operators defined as
\begin{equation}
P_{R}=\frac{1}{n!}\sum_{\sigma\in S_n}\chi_{R}(\sigma)\,\sigma,\label{Proj}
\end{equation}
that satisfy
\begin{equation}
P_{R}\,\sigma=\sigma\, P_{R},\qquad P_{R}P_{S}=\frac{\delta_{RS}}{d_R}P_{S},\qquad \Tr(P_R)=Dim (R)\label{properties}.
\end{equation}
In the above formulas $R$ and $S$ denote arbitrary Young diagrams, $\sigma$ the elements of the permutation group $S_n$ and $\chi_R(\sigma)$ stand for characters of $\sigma$ in representation $R$.
Below we use the projectors to present an elegant derivation of \eqref{2ptSchurs} obtained in \cite{Dey:2011ea}.\\
As demonstrated in \cite{Dey:2011ea}, Schurs in ABJ(M) can be written as
\begin{equation}
\chi_{R}(AB^{\dagger})=\delta_{RS}\frac{d_R}{(n!)^2}\sum_{\sigma,\rho\in S_{n}}\chi_{R}(\sigma)\chi_{S}(\rho)\Tr\left((\sigma A^{\otimes n})(\rho \left(B^{\dagger}\right)^{\otimes n})\right).
\end{equation}
This can be expressed in an elegant way in terms of \eqref{Proj} as
\begin{equation}
\chi_{R}(AB^{\dagger})=\delta_{RS}\,d_{R}\,\Tr(P_{R}A^{\otimes n}P_{S}\left(B^{\dagger}\right)^{\otimes n}).
\end{equation}
The two point function of Schurs after contracting $A$s and $B$s becomes
\begin{eqnarray}
\langle \chi_{R_1}(AB^{\dagger})\chi_{R_2}(A^{\dagger}B) \rangle&=& d_{R_1}d_{R_2}\sum_{\sigma,\rho \,\in \,S_n}\Tr\left(P_{R_1}\rho P_{R_2}\sigma^{-1}\right)\Tr\left(P_{R_2}\sigma P_{R_1}\rho^{-1}\right)\nonumber\\
&=&\delta_{R_1R_2}\,n!\sum_{\tilde{\rho}\,\in\,S_n}\Tr(P_{R_1}\tilde{\rho})\Tr(P_{R_2}\tilde{\rho}^{-1}).
\end{eqnarray}
where in the first line $\sigma$ and $\rho$ represent possible contractions and in the second equality we used \eqref{properties}, substituted
\begin{equation}
\tilde{\rho}=\rho \sigma^{-1},
\end{equation}
and summed over $\sigma$ what yield factor $n!$. 
Finally expanding traces of permutation in rems of characters and using the fundamental orthogonality relation for the sums over $\sigma_i$ and $\tilde{\rho}$ yields
\begin{equation}
\delta_{R_1R_2}\sum_{T_1T_2}\delta_{R_1T_1}\delta_{R_2T_2}\delta_{T_1T_2}\frac{n!Dim(T_1)n!Dim(T_2)}{d_{T_1}d_{T_2}}.
\end{equation}
which after inserting
\begin{equation}
Dim(R)=\frac{f_{R}(N)}{hooks_{R}},\qquad d_{R}=\frac{n!}{hooks_{R}},
\end{equation}
proves \eqref{2ptSchurs}.

\section{Open string correlators} \label{Details}
Let us start with the two point correlator of open strings. Notice that there are two possible correlators that we can write in ABJ theory. Ones in terms of $(AB^{\dagger})_{N\times N}$ and others in terms of $(A^{\dagger}B)_{M\times M}$. We call the former N-strings and the later M-strings. From the conservation of $U(N)(U(M))$ charge they are both constrained to the form
\begin{eqnarray}
\langle\left((AB^{\dagger})^J\right)^{i}_{j}\left((BA^{\dagger})^J\right)^{l}_{k}\rangle&=&C_1\delta^{i}_{j}\delta^{l}_{k}+C_2\delta^{i}_{k}\delta^{l}_{j}\label{first}\\
\langle\left((A^{\dagger}B)^J\right)^{\alpha}_{\beta}\left((B^{\dagger}A)^J\right)^{\lambda}_{\kappa}\rangle&=&D_1\delta^{\alpha}_{\beta}\delta^{\lambda}_{\kappa}+D_2\delta^{\alpha}_{\kappa}\delta^{\lambda}_{\beta},\label{second}
\end{eqnarray}
where we distinguish latin indices $i,j=1,...,N$ from greek $\alpha,\beta=1,...,M$, and constants $C_1, C_2$ and $D_1, D_2$ are to be determined. We begin with the correlator of N-strings \eqref{first}. There are two ways to contract the indices hence we have
\begin{eqnarray}
\langle\Tr((AB^{\dagger})^J)\Tr((BA^{\dagger})^J)\rangle&=&C_1\,N^2+C_2\,N\\
\langle\Tr((AB^{\dagger})^J(BA^{\dagger})^J)\rangle&=&C_1\,N+C_2\,N^2
\end{eqnarray}
To the leading order in $N(M)$ the second correlator can be expressed in terms of the two point function of the single trace operators. Notice that to the leading order in $N$ and $M$, only contractions between paris  $AB^{\dagger}$ matter. This is because
\begin{equation}
(\check{A}\check{B}^{\dagger})^{i}_{j}(B^{\dagger}A)^{l}_{k}=\check{A}^{i}_{\alpha}(\check{B}^{\dagger})^{\alpha}_{j}(B^{\dagger})^{l}_{\beta}A^{\beta}_{k}=\delta^{i}_{k}\delta^{\beta}_{\alpha}\delta^{\alpha}_{\beta}\delta^{l}_{j}
=M\,\delta^{i}_{k}\delta^{l}_{j}
\end{equation}
Hence if we perform just a single contraction in the two point correlator of $J+1$ pairs, the leading answer is
\begin{equation}
\langle\Tr((AB^{\dagger})^{J+1})\Tr((BA^{\dagger})^{J+1})\rangle\sim M(J+1)\langle\Tr((AB^{\dagger})^J(BA^{\dagger})^J)\rangle.
\end{equation}
Finally we arrive at two equations
\begin{eqnarray}
\langle J\rangle\equiv\langle\Tr((AB^{\dagger})^J)\Tr((BA^{\dagger})^J)\rangle&=&C_1\,N^2+C_2\,N\\
\langle J+1\rangle\equiv\langle\Tr((AB^{\dagger})^{J+1})\Tr((BA^{\dagger})^{J+1})\rangle&\sim&M(J+1)\left(C_1\,N+C_2\,N^2\right),
\end{eqnarray}
that are solved to
\begin{eqnarray}
C_1&\sim&\frac{1}{N^3-N}\left(N\,\langle J\rangle-\frac{\langle J+1\rangle}{M(J+1)}\right)\sim (J-1)M^{J}N^{J-2}\\
C_2&\sim&\frac{1}{N^3-N}\left(\frac{N\langle J+1\rangle}{M(J+1)}-\langle J\rangle\right)\sim M^{J}N^{J-1}
\end{eqnarray}

\medskip
The three point functions are important for studying string dynamics such as splitting and joining of open strings, which is a known phenomenon in the case of giant graviton. The three point function of this type is
\begin{align}
\langle((AB^{\dagger})^{J_{1}})_{i}^{j}((AB^{\dagger})^{J_{2}})_{k}^{l}((BA^{\dagger})^{J_{1}+J_{2}})_{p}^{q}\rangle=&\delta_{i}^{j}\delta_{k}^{l}\delta_{p}^{q}C_{1}+\delta_{i}^{l}\delta_{k}^{j}\delta_{p}^{q}C_{2}+\delta_{i}^{j}\delta_{k}^{q}\delta_{p}^{l}C_{3}\nonumber\\
&+\delta_{i}^{q}\delta_{k}^{l}\delta_{p}^{j}C_{4}+\delta_{i}^{l}\delta_{k}^{q}\delta_{p}^{j}C_{5}+\delta_{i}^{q}\delta_{k}^{j}\delta_{p}^{l}C_{6}
\end{align}
There are six possible contractions which lead to the following three point functions
\begin{align}
A\equiv\langle\textrm{Tr}((AB^{\dagger})^{J_{1}})\textrm{Tr}((AB^{\dagger})^{J_{2}})\textrm{Tr}((BA^{\dagger})^{J_{1}+J_{2}})\rangle=&N^{3}C_{1}+N^{2}(C_{2}+C_{3}+C_{4})\nonumber\\
&+N(C_{5}+C_{6}),\nonumber\\
B\equiv\langle\textrm{Tr}((AB^{\dagger})^{J_{1}+J_{2}})\textrm{Tr}((BA^{\dagger})^{J_{1}+J_{2}})\rangle=&N^{3}C_{2}+N^{2}(C_{1}+C_{5}+C_{6})\nonumber\\
&+N(C_{3}+C_{4}),\nonumber\\
C\equiv\langle\textrm{Tr}((AB^{\dagger})^{J_{1}})\textrm{Tr}((AB^{\dagger})^{J_{2}}(BA^{\dagger})^{J_{1}+J_{2}})\rangle=&N^{3}C_{3}+N^{2}(C_{1}+C_{5}+C_{6})\nonumber\\
&+N(C_{2}+C_{4}),\nonumber\\
D\equiv\langle\textrm{Tr}((AB^{\dagger})^{J_{2}})\textrm{Tr}((AB^{\dagger})^{J_{1}}(BA^{\dagger})^{J_{1}+J_{2}})\rangle=&N^{3}C_{4}+N^{2}(C_{1}+C_{5}+C_{6})\nonumber\\
&+N(C_{2}+C_{3}),\nonumber\\
E\equiv\langle\textrm{Tr}((AB^{\dagger})^{J_{1}+J_{2}}(BA^{\dagger})^{J_{1}+J_{2}})\rangle=&N^{3}C_{5}+N^{2}(C_{2}+C_{3}+C_{4})\nonumber\\
&+N(C_{1}+C_{6}),\nonumber\\
F\equiv\langle\textrm{Tr}((AB^{\dagger})^{J_{1}+J_{2}}(BA^{\dagger})^{J_{1}+J_{2}})\rangle=&N^{3}C_{6}+N^{2}(C_{2}+C_{3}+C_{4})\nonumber\\
&+N(C_{1}+C_{5}).\nonumber
\end{align}
It is easy to see that in the above set {$E=F$}. Now solving these equations, we find
\begin{align}
C_{1}=&\frac{1}{(N^{2}-1)(N^{2}-4)}\Bigl[\frac{N^{2}-2}{N}A-B-C-D+\frac{4}{N}E\Bigr],\nonumber\\
C_{2}=&\frac{1}{(N^{2}-1)(N^{2}-4)}\Bigl[-A+\frac{N^{2}-2}{N}B+\frac{2}{N}C+\frac{2}{N}D-2E\Bigr],\nonumber\\
C_{3}=&\frac{1}{(N^{2}-1)(N^{2}-4)}\Bigl[-A+\frac{2}{N}B+\frac{N^{2}-2}{N}C+\frac{2}{N}D-2E\Bigr],\nonumber\\
C_{4}=&\frac{1}{(N^{2}-1)(N^{2}-4)}\Bigl[-A+\frac{2}{N}B+\frac{2}{N}C+\frac{N^{2}-2}{N}D-2E\Bigr],\nonumber\\
C_{5}=&C_{6}=\frac{1}{(N^{2}-1)(N^{2}-4)}\Bigl[\frac{2}{N}A-B-C-D+NE\Bigr].
\end{align}\\

We note that, there are only two new correlators {$C$} and {$D$} that need to be computed. The other correlators can directly be computed with advent of section 4. To find the new correlators, we consider the leading terms of the three point function. That is
\begin{align}
\langle\textrm{Tr}((AB^{\dagger})^{J_{1}})\textrm{Tr}((AB^{\dagger})^{J_{2}})\textrm{Tr}((BA^{\dagger})^{J_{1}+J_{2}})\rangle\sim&MJ_{1}\langle\textrm{Tr}((AB^{\dagger})^{J_{2}})\textrm{Tr}((AB^{\dagger})^{J_{1}-1}(BA^{\dagger})^{J_{1}+J_{2}-1})\rangle\nonumber\\
+&MJ_{2}\langle\textrm{Tr}((AB^{\dagger})^{J_{1}})\textrm{Tr}((AB^{\dagger})^{J_{2}-1}(BA^{\dagger})^{J_{1}+J_{2}-1})\rangle\nonumber
\end{align}
Upon using equation \eqref{leading-terms-three-point}, we get
\begin{equation*}
\langle\textrm{Tr}((AB^{\dagger})^{J_{2}})\textrm{Tr}((AB^{\dagger})^{J_{1}-1}(BA^{\dagger})^{J_{1}+J_{2}-1})\rangle=J_{1}J_{2}N^{J_{1}+J_{2}-1}M^{J_{1}+J_{2}-1}+J_{1}J_{2}M^{J_{1}+J_{2}}N^{J_{1}+J_{2}}
\end{equation*}
Shifting $J_1$ to $J_1+1$, we arrive at
\begin{equation}
\langle\textrm{Tr}((AB^{\dagger})^{J_{2}})\textrm{Tr}((AB^{\dagger})^{J_{1}}(BA^{\dagger})^{J_{1}+J_{2}})\rangle=J_{2}(J_{1}+1)\left(N^{J_{1}+J_{2}}M^{J_{1}+J_{2}}+M^{J_{1}+J_{2}+1}N^{J_{1}+J_{2}+1}\right)
\end{equation}
Similarly
\begin{equation}
\langle\textrm{Tr}((AB^{\dagger})^{J_{1}})\textrm{Tr}((AB^{\dagger})^{J_{2}}(BA^{\dagger})^{J_{1}+J_{2}})\rangle=J_{1}(J_{2}+1)\left(N^{J_{1}+J_{2}}M^{J_{1}+J_{2}}+M^{J_{1}+J_{2}+1}N^{J_{1}+J_{2}+1}\right)
\end{equation}
Recall the result of open string correlators in {${\cal{N}}=4$} SYM theory \cite{de Mello Koch:2007uu}
\begin{equation}
\left\langle \textrm{Tr}(Y^{J_{1}})\textrm{Tr}(Y^{J_{2}}(Y^{\dagger})^{J_{1}+J_{2}})\right\rangle =J_{1}(J_{2}+1)N^{J_{1}+J_{2}}
\end{equation}
once again, ABJ answer is not simply a product of two open string correlators. The solution for $C_1$, $C_2$, $C_3$, $C_4$ and $C_5$ in the large $N$, $M$ limit is
\begin{align}
C_{1}=&[J_{1}J_{2}(J_{1}+J_{2})+4]M^{J_{1}+J_{2}}N^{J_{1}+J_{2}-4}-2J_{1}J_{2}M^{J_{1}+J_{2}+1}N^{J_{1}+J_{2}-3},\nonumber\\
C_{2}=&(J_{1}+J_{2}-2)M^{J_{1}+J_{2}}N^{J_{1}+J_{2}-3},\nonumber\\
C_{3}=&(J_{1}J_{2}+J_{1})M^{J_{1}+J_{2}+1}N^{J_{1}+J_{2}-2}-2M^{J_{1}+J_{2}}N^{J_{1}+J_{2}-3},\nonumber\\
C_{4}=&(J_{1}J_{2}+J_{2})M^{J_{1}+J_{2}+1}N^{J_{1}+J_{2}-2}-2M^{J_{1}+J_{2}}N^{J_{1}+J_{2}-3},\nonumber\\
C_{5}=&C_{6}=M^{J_{1}+J_{2}}N^{J_{1}+J_{2}-2}.
\end{align}
It is now easy to compute the amplitude of closed string propagating between two excited D-brane states, since we need to compute the leading contribution of the correlator $\bigl\langle\textrm{Tr}(AB^{\dagger J_{1}})(AB^{\dagger J_{2}})_{j}^{i}(BA^{\dagger(J_{1}+J_{2})})_{k}^{l}\bigr\rangle$. From the above result, we get
\begin{equation}
\bigl\langle\textrm{Tr}(AB^{\dagger J_{1}})(AB^{\dagger J_{2}})_{j}^{i}(BA^{\dagger(J_{1}+J_{2})})_{k}^{l}\bigr\rangle=M^{J_{1}+J_{2}}N^{J_{1}+J_{2}-2}\delta_{k}^{i}\delta_{j}^{l}.
\end{equation}

\end{appendix}



\end{document}